\documentclass[epj]{svjour}

\usepackage[utf8]{inputenc}
\usepackage{graphicx}
\usepackage{flafter}

\usepackage{amsmath,amssymb,amstext,amsfonts}
\usepackage{braket}
\usepackage{bm}
\usepackage{float}
\usepackage{color}
\usepackage{subfigure}
\usepackage{blindtext}
\usepackage{marginnote}

\newcommand{\hf}[1]{\textcolor{black}{#1}}

\begin{document}
	
	\title{Electronic Properties of $\alpha-\mathcal{T}_3$ Quantum Dots in Magnetic Fields}

	\titlerunning{$\alpha-\mathcal{T}_3$ quantum dot}
	
	\author{Alexander Filusch\thanks{alexander.filusch@uni-greifswald.de} \and Holger Fehske\thanks{fehske@physik.uni-greifswald.de}}
	
	\authorrunning{A. Filusch et al.}
	

	\institute{%
		Institute of Physics, University Greifswald, 17487 Greifswald, Germany}


	\abstract{
		We address the electronic properties of quantum dots in the two-dimensional  $\alpha-\mathcal{T}_3$ lattice when subjected to a perpendicular magnetic field.  Implementing an infinite mass boundary condition, we first solve the eigenvalue problem  for an isolated quantum dot  in the low-energy, long-wavelength  approximation where the system is described by an effective Dirac-like Hamiltonian that interpolates between the graphene (pseudospin 1/2) and Dice (pseudospin 1) limits. Results are compared  to a full numerical (finite-mass)  tight-binding lattice calculation. In a second step we analyse charge transport through a contacted  $\alpha-\mathcal{T}_3$  quantum dot in a magnetic field by calculating the local density of states and the conductance  within the kernel polynomial and Landauer-B\"uttiker approaches.   Thereby the influence of a disordered environment is discussed as well.}
	
	\maketitle   
	\section{Introduction} 
Quantum matter with Dirac-cone functionality is expected 	to provide  the building block of future electronics, plasmonics and photonics.   Against this background, above all graphene-based nanostructures were intensively examined, both experimentally and theoretically, in the recent past. This is because their striking electronic properties  can be modified by nanostructuring and patterning, e.g., manufacturing nanoribbons~\cite{YPSCL07}, nanorings~\cite{ROWHSVM08}, junctions~\cite{WDM07}, quantum dots~\cite{PSKYHNG08}, or  even quantum dot arrays~\cite{VAW11,CCOWK16}. Thereby the transport behaviour heavily relies on the geometry of the sample (or device) and its edge shape~\cite{GALMW07,ZG09}.  

The mutability of systems with Dirac nodal points, which is especially important from a technological point of view~\cite{Guea12},  can also be achieved by applying external  electric  (static or time-dependent) fields. One of the options are nanoscale top gates that modify the electronic structure  in a restricted area~\cite{GMSHG08}. This allows to imprint junctions and barriers relatively easy, and therefore opens new possibilities to study fascinating phenomena such as Klein tunnelling~\cite{Kl28,KNG06}, Zitterbewegung~\cite{TBM07,WF18}, particle confinement~\cite{BTB09,FHP15}, Veselago lensing~\cite{Brea19},  Mie scattering analogues~\cite{HBF12b,CPP07,HBF13a,PHF13,AU14} and resonant scattering~\cite{WF14,FWF20}. Clearly the energy of the charge-carrier states can be manipulated by (perpendicular) magnetic fields as well. With this  the quantum Hall effect, the Berry phase curvature,  the Landau level splitting and Aharonov-Bohm oscillations have been investigated~\cite{ROWHSVM08,ZTSK05,SSH12}.

Shortly after the field of graphene was opened, Dirac-cone physics was combined with flat-band physics in a modified lattice, the  $\alpha-\mathcal{T}_3$ lattice, which is obtained by coupling one of the inequivalent sites of the honeycomb lattice to an additional atom located at the centre of the hexagons with strength $\alpha$~\cite{Su86,VMD98,DKM11}. Obviously, such a lattice interpolates between graphene ($\alpha=0$) and the Dice lattice  ($\alpha=1$). Most notably, the flat band crosses the nodal Dirac points, which has peculiar consequences, such as an $\alpha$-dependent Berry phase~\cite{RMFPM14}, super-Klein tunnelling~\cite{SSWX10,UBWH11}, or Weiss oscillations~\cite{FD17}. \hf{Interestingly, the magneto-optical response will be also enhanced due to the flat bands~\cite{CXWLM19}. Analysing the frequency-dependent magneto-optical and zero-field conductivity of $\rm Hg_{1-x}Cd_xTe$~\cite{Orea14} at the critical cadmium concentration $x_c\simeq 0.17$ (marking the semimetal-semiconductor transition), it has been shown that this material can be linked to the $\alpha-\mathcal{T}_3$  model with $\alpha=1/\sqrt{3}$~\cite{MN15}. Other possibilities to realise the $\alpha-\mathcal{T}_3$ and Dice ($\alpha=1$) models  experimentally are cold bosonic or fermionic atoms loaded in optical lattices~\cite{RMFPM14,RCF06}.} 

The massless Dirac equation~\cite{Di28} provides the basis for numerous theoretical investigations of the low-energy excitations in these novel, strictly two-dimensional systems~\cite{SESI08,JMPT09,BUGH09,DKM11,UBWH11,VOVSDCD13,BCGC17,XL16,SZ16,IN17,TP17,ZC17,WXHL17,BO19,HIXLG19,XL20a,XL20b}, whereby the quasiparticles carry a pseudospin~1 in the  Dice lattice rather than pseudospin~1/2 in the case of graphene.  Accordingly one usually works with a three- (Dice) and two-component (graphene) realisation of the standard Dirac-Weyl Hamiltonian. Investigating the electronic properties of $\alpha-\mathcal{T}_3$ quantum dots in magnetic fields, we also start from such a description, and therefore must implement a boundary condition when the dot is cut out from the plane~\cite{AB08,SESI08,GZCTFP11}. Of course, this approach has to be approved by  comparison with lattice model results obtained numerically~\cite{WWBR10,GZCTFP11,FWPF18}. Addressing the transport behaviour of contacted dots and the influence of disorder on that we have to work with the full lattice model in any case.

The outline of this paper is as follows. In Section~\ref{theo} we introduce the  $\alpha-\mathcal{T}_3$ model, discuss the continuum approach, derive the infinite-mass boundary condition, and solve the eigenvalue problem for an isolated quantum dot in a constant 
magnetic field in dependence on $\alpha$. Section~\ref{results} contains our numerical results for the eigenvalue spectrum, the (local) density of states and the conductance.  Thereby we critically examine how the continuum model results compare to the numerical exact tight-binding lattice-model data (Section ~\ref{iso-dot}). Afterwards we study transport through a quantum dot subject to a magnetic field 
in the end-contacted lead-sample geometry most relevant for experiments (Section ~\ref{con-dot}), and analyse boundary disorder effects  (Section~\ref{dis-dot}). We conclude in Section.~\ref{conclusions}.

	\section{Theoretical approach} 
	\label{theo}
	\subsection{$\alpha-\mathcal{T}_3$ model}  
\begin{figure}[t]
		\begin{minipage}[c]{0.44\textwidth}
			\hspace*{0.2cm}
			\includegraphics[width=1\linewidth]{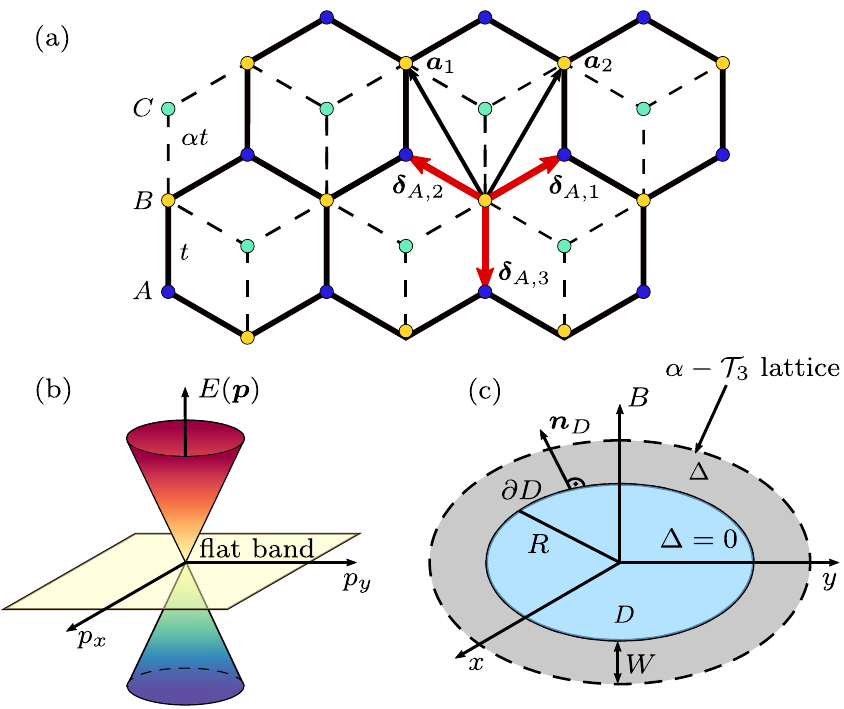}\\
		\end{minipage}
		\caption{(a) $\alpha-\mathcal{T}_3$ lattice with basis $\{A,B,C\}$  and Bravais-lattice vectors $\mathbf{a}_1$ and $\mathbf{a}_2$. Next-nearest neighbours are connected by $\bm{\delta}_{A,i}$ ($i=1,2,3$) where $\alpha$ gives the ratio of the transfer amplitudes $A$-$B$ and $B$-$C$. In the numerical work we use graphene-like parameters, i.e., a lattice constant $a=0.142$~nm and a transfer integral $t=3.033$~eV which sets the energy scale. (b) Continuum model energy dispersion near $K$ or $K'$ when $\Delta=0$ with two linear dispersive bands and a flat band at $E=0$. (c) $\alpha-\mathcal{T}_3$ dot setup with a constant magnetic field, perpendicular to the $(x,y)$ plane. The quantum dot $D$ (blue region) with radius $R$ and zero gap ($\Delta=0$) is surrounded by a ring of width $W$ (grey, dashed border) having a gapful band structure ($\Delta>0$). The vector $\mathbf{n}_D$ is perpendicular to the boundary.  }
		\label{fig1}
	\end{figure}

	We start from the  tight-binding Hamiltonian  
	\begin{align} 
	H^\alpha  =& - \sum \limits_{\langle i j\rangle}te^{i\Phi_{ij}} a^\dagger_i b_j^{}- \sum\limits_{\langle i j\rangle}\alpha te^{i\Phi_{ij}}  b^\dagger_i c_j^{}  \nonumber\\&+ \Delta \sum \limits_i\left( a_i^\dagger a_i^{} -b_i^\dagger b_i^{} +c^\dagger_i c_i^{}\right) + \text{H.c.}\,, \label{eq:Tight-Binding}
	\end{align}   
	where $a^{(\dagger)}$, $b^{(\dagger)}$ and $c^{(\dagger)}$ annihilate (create) a particle in a Wannier state centred	 at site $A$, $B$ and $C$ of the $\alpha-\mathcal{T}_3$ lattice, respectively. The nearest-neighbour transfer amplitude between $A$ and $B$ sites is given by $t$, and  will be rescaled by $\alpha$ if hopping takes place between nearest-neighbour $B$ and $C$ sites, see Fig.~\ref{fig1} (a). In this way, the scaling parameter interpolates between the honeycomb lattice ($\alpha=0$) and the Dice lattice ($\alpha=1$). In the presence of a vector potential  $\mathbf{A}(\mathbf{r})$, hopping is modified further by the Peierls phase $\Phi_{ij}=2\pi/\phi_0\int_i^j \mathbf{A}(\mathbf{r}) \mathrm{d}\mathbf{r}$ with  $\phi_0=h/e$.  
	
	In order to implement boundary conditions below,  we have introduced  a  sublattice-dependent onsite potential  $\Delta$, which opens a gap in the band structure  at the charge neutrality point. In what follows we assume that $\Delta>0$; the case $\Delta<0$ is obtained by changing the sign of the energy~$E$.  Note that a positive $\Delta$ will shift the flat band to the bottom of the upper dispersive one. 
 	
	Next we write down the corresponding continuum Dirac-Weyl Hamiltonian  in momentum space in the absence of a magnetic field, being valid for low energies near the Dirac-points $K$ ($\tau=+1$) and $K'$ ($\tau=-1$):
	\begin{align}
	H^\varphi_\tau = v_\mathrm{F} \mathbf{S}^\varphi_\tau \cdot \mathbf{p} +  U \Delta \,,
	\label{eq:DW-QT}
	\end{align}
	where $\varphi= \arctan \alpha$ and $\tau$ is the valley index.  
	In equation~\eqref{eq:DW-QT}, $v_\mathrm{F}=3at/2\hbar$ is the Fermi velocity, where $a$ refers to the lattice constant, and $\mathbf{p}=-i\hbar\bm{\nabla}$ denotes the momentum operator in two spatial dimensions.  The components of the pseudospin vector $\mathbf{S}_\tau^\varphi=(\tau S_x^\varphi, S_y^\varphi)$ in (three-dimensional) spin space, 
	\begin{align}
	S_x^\varphi&=\begin{pmatrix}
	0 & \cos\varphi & 0 \\
	\cos\varphi & 0 & \sin\varphi \\
	0 & \sin\varphi	& 0 
	\end{pmatrix}, \nonumber\\[0.1cm]
	S_y^\varphi &= \begin{pmatrix}
	0 & -i\cos\varphi & 0 \\
	i\cos\varphi & 0 & -i\sin\varphi \\
	0 & i\sin\varphi	& 0 
	\end{pmatrix} \,,
	\label{eq:ps-matrices}
	\end{align}
	represent the sublattice degrees of freedom. In~equation~(\ref{eq:DW-QT}), the  matrix 
	\begin{align}
	U = \begin{pmatrix}
	1 & 0 & 0 \\ 
	0 & -1 & 0 \\
	0 & 0 & 1
	\end{pmatrix}
	\end{align}
	introduces a mass term, similar to $\sigma_z$ in the standard (spin-1/2) massive Dirac-Weyl equation. Therefore $H^\varphi_\tau$ comprises 
	the limiting cases of massive pseudospin 1/2 ($\alpha=0$)  and pseudospin 1 ($\alpha=1$) Dirac-Weyl quasiparticles. Rescaling the energy by $\cos\varphi$,
	the eigenvalues  $E_{\tau, s} |\psi_\tau\rangle$  of $H^\varphi_\tau |\psi_\tau \rangle = E_{\tau, s} |\psi_\tau \rangle$ become
	\begin{align}\label{eq:eigenvalues1}
	E_{\tau,0} &= \Delta\,,\\
	E_{\tau,s}  &= s \sqrt{(v_\mathrm{F}\mathbf{p})^2+\Delta^2},
	\label{eq:eigenvalues}
	\end{align}
	where $s=\pm 1$ marks the band index. Note that the energy eigenvalues are valley degenerate.

	\subsection{Infinite mass boundary condition}
	Implementing the so-called infinite mass boundary condition (IMBC) we take up a proposal by Berry and Mondragon~\cite{BM87}. 
	For this, we consider the  Hamiltonian 
	\begin{align}
	H^\varphi_\tau= \mathbf{S}^\varphi_\tau \cdot \mathbf{p} + \Delta(\mathbf{r}) U 
	\label{eq:H-mass-RB}
	\end{align}
	(setting $v_\mathrm{F}=\hbar=1$ in this section), with a position-dependent mass term, $\Delta(\mathbf{r}) U$, which  is zero (finite) inside (outside) a circular region $D$, cf. Fig. \ref{fig1}. Note that Hermiticity of the Hamiltonian in $D$ implies $\langle \mathbf{n}_{D} \cdot \mathbf{j}^\tau \rangle(\mathbf{r}) = 0$ at every point $\mathbf{r}$ of the boundary $\partial D$.  Here, $\mathbf{j}^\tau=\mathbf{S}^\varphi_\tau$ is the current density operator and $\mathbf{n}_{D}=(\cos\vartheta(\mathbf{r}),\sin\vartheta(\mathbf{r}))$ is the normal vector of  $D$. Then the  local boundary condition for a general wave function $\psi_\tau = (\psi_{\tau,A},\, \psi_{\tau,B},\, \psi_{\tau,C})$ is
	\begin{align}
	\psi_{\tau,B}\bigg|_{\mathbf{r}\in \partial D}= i \Gamma_\tau (\mathbf{r})&\left(\cos\varphi \;e^{i\tau\vartheta(\mathbf{r})}\psi_{\tau,A} \nonumber\right.\\&\left.+\sin\varphi \;e^{-i\tau\vartheta(\mathbf{r})}\psi_{\tau,C}\right)  \bigg|_{\mathbf{r}\in \partial D}\, .\label{Masserandbedingung}
	\end{align}
		
	The variable $\Gamma_\tau (\mathbf{r})$ can be obtained from the solution of the scattering problem at a planar  mass step, 
	$H^\varphi_\tau= \mathbf{S}_\tau^\varphi \cdot \mathbf{p}+ \Delta\Theta(x)$, where the height of the barrier is assumed to be larger than the energy ($\Delta >|E|$) and the Heaviside step function divides the $(x,y)$-plane in regions~I for $x<0$ and~II for $x>0$. 
	In doing so, we will consider only the dispersive states, since $\langle\mathbf{j^\tau}\rangle = 0$  for the flat band states.
	
	In region~I, the wave function with wave vector $\mathbf{k}=(k_x, k_y)$ and propagation direction $\theta_\mathbf{k}=\arctan k_y/k_x$  is 
	\begin{align}
	\psi^{\rm I}_{\tau,s} = \frac{1}{\sqrt{2}}& \begin{pmatrix}
	\tau\cos\varphi \;e^{-i\tau\theta_{\mathbf{k}}} \\
	s \\
	\tau\sin\varphi \;e^{i\tau\theta_{\mathbf{k}}}
	\end{pmatrix} e^{i\mathbf{k} \mathbf{r}} \nonumber \\& + \frac{r_\tau}{\sqrt{2}} \begin{pmatrix}
	\tau\cos\varphi \;e^{i\tau\theta_{\mathbf{k}}} \\
	-s\\
	\tau\sin\varphi \;e^{-i\tau\theta_{\mathbf{k}}}
	\end{pmatrix} e^{i\mathbf{k'} \mathbf{r}}\,. \label{eq:wvf-(i)}
	\end{align}
	Here, $\mathbf{k}'= (-k_y, k_x)$ denotes the wave vector of the reflected wave having a valley-dependent reflection coefficient $r_\tau$.
	
	In region~II, the wave function takes the form
	\begin{align}
	\psi_{\tau,s}^{\rm II}=\frac{t_{\tau}}{\sqrt{2}}\begin{pmatrix}
	\tau a_{\tau,s}\\
	b_{\tau,s}\\
	\tau b_{\tau,s}\\
	\end{pmatrix}\frac{e^{-qx+ik_y y}}{d_{\tau,s}}\,, \label{eq:trans_u}
	\end{align}
	where $t_\tau$ denotes  the valley-dependent transmission coefficient, $(k_x,k_y)=(iq,k_y)$, and
	\begin{align}
	a_{\tau,s} &= - i \cos\varphi \sqrt{(q-\tau k_y)^2(\Delta+ E)},\\
	b_{\tau,s} &= \sqrt{(q^2-k_y^2)(\Delta-E)}, \\
	c_{\tau,s} &= - i \sin\varphi \sqrt{(q+\tau k_y)^2(\Delta+ E)}, \\
	d_{\tau,s} &= \sqrt{\Delta q^2+ Ek_y^2-\tau k_y q\cos 2\varphi(\Delta+ E)}.
	\end{align}
	Obviously,  $\psi_{\tau,s}^{\rm II}$ is an evanescent wave perpendicular to the boundary  but oscillatory along $\partial D$. 
	
	Enforcing the continuity of the wave function at $x=0$, 
	\begin{align}
	\psi_{\tau,s,B}^{\rm I} &= \psi_{\tau,s,B}^{\rm II}, \\
	\cos\varphi\, \psi_{\tau,s,A}^{\rm I}+\sin\varphi \, \psi_{\tau,s,C}^{\rm I} &=\cos\varphi\, \psi_{\tau,s,A}^{\rm II}\nonumber\\&\quad+\sin\varphi \, \psi_{\tau,s,C}^{\rm II}\,, \label{eq:wvf-continuity}
	\end{align}
	and performing the limit $\Delta\rightarrow\infty$ ($q\rightarrow\infty$), we obtain
	\begin{align}
	r_{\tau,s}=\frac{ is+\cos^{2}\varphi \;e^{-i\tau\theta_{\mathbf{k}}}+\sin^{2}\varphi \;e^{+i\tau\theta_{\mathbf{k}}}}{ is -\cos^{2}\varphi \;e^{i\tau\theta_{\mathbf{k}}}+\sin^{2}\varphi \;e^{-i\tau\theta_{\mathbf{k}}}} \label{eq:r_u}.
	\end{align}
	Since $|r_{\tau,s}|^2=1$ $\forall E$, the incoming wave is perfectly reflected at the boundary, regardless of $\tau$ and $s$. Inserting the full wave function~\eqref{eq:wvf-(i)} with~\eqref{eq:r_u} and $\mathbf{n}_{D}(x=0)\equiv\mathbf{e}_x$ into equation ~\eqref{Masserandbedingung}, we find  $\Gamma_\tau =  \tau$.
	
	Clearly the whole scattering problem can be rotated by any angle $\vartheta$, i.e., for the $\alpha-\mathcal{T}_3$ lattice  the IMBC  at $\partial D$ becomes:
	\begin{align}
	\psi_{\tau,B}=  i \tau \left(\cos\varphi\psi_{\tau,A} \;e^{i\tau\vartheta}+\sin\varphi\psi_{\tau,C} \;e^{-i\tau\vartheta}\right). \label{eq:IMBC}
	\end{align}
	At $\alpha=0$ we reproduce the IMBC of graphene~\cite{BM87}.

	\subsection{Eigenvalue problem of the  $\alpha-\mathcal{T}_3$  quantum dot in a perpendicular magnetic field} We now consider a circular quantum dot of radius $R$ in a constant magnetic field, $\mathbf{B}=B\mathbf{e}_z$,  related to the vector potential $\mathbf{A}= B/2(-y,x,0)$. Then, using polar coordinates  $(x,y)\rightarrow (r,\phi)$, the (minimal-coupling) Hamiltonian is
	\begin{align}
	H^\varphi_\tau = v_\mathrm{F} \mathbf{S}^\varphi_\tau \cdot (\mathbf{p}+e\mathbf{A}) + \Delta U\Theta(r-R)\,. \label{eq:H-dot-cont}
	\end{align}
	In the quantum dot region $D$ ($r<R$) we have $\Delta=0$ and 
	\begin{align}
	H^\varphi_\tau= \tau \hbar \omega_{c}\begin{pmatrix}
	0 &\cos\varphi L_{\tau, -} & 0 \\
	\cos \varphi L_{\tau, +} & 0 & \sin\varphi L_{\tau, -} \\
	0 & \sin\varphi L_{\tau, +} & 0 \\
	\end{pmatrix}\,. \label{eq:H_Bfeld}
	\end{align}
	Here, $L_{\tau, \mp} = -i e^{\mp i\tau \phi}\left\{\partial_\rho\pm \frac{\tau L_\mathrm{z}}{\hbar\rho} \pm \tau\rho \right\}$, $L_\mathrm{z} = -i\hbar \partial_\phi$, 	$\hbar \omega_{c}=\sqrt{2}\hbar v_\mathrm{F}/l_{B}$, $l_{B}=\sqrt{\frac{\hbar}{e B}}$, and $\rho=r/\sqrt{2}l_{B}$. Rotational symmetry  ($[H^\varphi_0, J_z ]= 0$ ) suggests the ansatz:
	\begin{align}
	\psi_{\tau}^{D}=\begin{pmatrix}
	\chi_{\tau,A}\, e^{i(m-\tau)\phi} \\
	\chi_{\tau,B}\,e^{i m \phi} \\
	\chi_{\tau,C}\, e^{i(m+\tau)\phi}\\
	\end{pmatrix}. \label{eq:Ansatz_kommutator}
	\end{align}

	With this, for the {\it dispersive band states},  we obtain the following differential equation for the $\chi_{\tau,B}$ component: 
	\begin{align}
	0 = &\left\{\partial^2_\rho +\frac{1}{\rho}\partial_\rho -2m +4\varepsilon_\tau^2 \right.\nonumber\\& \left.\;\;\;+2\tau\cos{2\varphi} -\left(\frac{m^2}{\rho^2}+	\rho^2\right)\right\}\chi_{\tau, B} \label{eq:DGL_chi_h},
	\end{align}
	yielding 
	\begin{align}
	\psi_{\tau,s}^{D} = N\begin{pmatrix}
	\tau\cos\varphi \rho^{-m+\tau} f_{\tau,A} e^{i(m-\tau)\phi} \\
	i\varepsilon_\tau \rho^{-m} L^{-m}_{n_\tau}(\rho^2) e^{im\phi}\\
	\tau\sin\varphi \rho^{-m-\tau} f_{\tau,C}e^{i(m+\tau)\phi}
	\end{pmatrix} \,e^{-\frac{\rho^2}{2}} \,.\label{eq:wavefunction_B}
	\end{align}
	The $L_a^b(x)$ are the generalized Laguerre polynomials, 
	\begin{align}
	f_{\tau,A} &=  \begin{cases}
	- L^{-m+1}_{n_+-1}(\rho^2) , &\text{ if } \tau=+1 \\
	(n_-+1) L^{-m-1}_{n_-+1}(\rho^2) , &\text{ if } \tau=-1\\[0.2cm]
	\end{cases}\\f_{\tau,C} &= f_{-\tau,A}\,,
	\end{align}
	 $n_\tau = \varepsilon_\tau^2 +(\tau\cos2\varphi-1)/2$ is the principal quantum number, $\varepsilon_\tau^2 = (E_{\tau,s}/\hbar \omega_\mathrm{c})^2$, $m$ is the total angular quantum number, and  $N$ is a normalization constant. 
	Note that $\chi_{\tau,C} \neq\chi_{-\tau,A}$  $\forall\varphi$, implying a valley asymmetry for \mbox{$\alpha<1$}.

	Employing now the IMBC~\eqref{eq:IMBC} for  $r=R$, where $\mathbf{n}_{D}=\mathbf{n}_{r}=(\cos\phi,\sin\phi)$, we obtain 
	\begin{align}
	0 = &\cos^2\varphi \rho^{2\tau}f_{\tau,A}(\rho) \nonumber\\&\quad+ \sin^{2}\varphi f_{\tau,C}(\rho)-\varepsilon_\tau \rho^\tau L_{n_\tau}^{-m}(\rho)\bigg|_{\rho=R/\sqrt{2}l_\mathrm{B}}. \label{eq:RB_dot}
	\end{align} 
	As a result, the energy eigenvalues $E_{\tau, s n_\rho, m,}$ are determined by the (positive and negative) zeros of this equation, where $n_\rho=1,2,3\ldots$ is the radial quantum number. At $\alpha=0$, these eigenvalues are related to those derived previously for graphene~\cite{TP17,SESI08,GZCTFP11} by replacing $m\to (m-1)$.

	In the large-$R$ (or large-$B$) limit, we can exploit the relation between Laguerre polynomials $L_a^b(x)$ and confluent hypergeometric functions of the first kind $M(a,b,x)$: 
	\begin{align}
	L_a^b(x) = \begin{pmatrix}
	a+b \\
	b
	\end{pmatrix} M(-a,b+1,x)\,. \label{eq:entwicklung1}
	\end{align}
	In leading order, $M(-a, b+1,x\to\infty)$  takes the form~\cite{AS70}:
	\begin{align}
	M(-a, b+1,x) = \frac{\Gamma(b+1)}{\Gamma(-a)}e^x x^{-a-b-1} \left[1+O(|x|^{-1})\right]. \label{eq:entwicklung2}
	\end{align}
	Substituting this into equation~\eqref{eq:RB_dot}, we obtain $\sin(\pi n_\tau)=0$. 
	Consequently $n_\tau =0$, 1, 2,\dots and the energy eigenvalues (Landau levels) become \cite{RMFPM14}
	\begin{align}
	E_{\tau,n_\tau,s} = s\hbar \omega_c \sqrt{n_\tau+\frac{1}{2}\left(1-\tau\cos2\varphi\right)}. \label{eq:LL}
	\end{align}
	
	For the {\it flat band states}, a similiar calculation gives
	\begin{align}
	\psi_{\tau, 0}^{D} = \begin{pmatrix}
	\sin\varphi \rho^{-m+\tau} f_{\tau,A} e^{i(m-\tau)\phi} \\
	0\\
	\cos\varphi \rho^{-m-\tau} f_{\tau,C} e^{i(m+\tau)\phi}
	\end{pmatrix}  e^{-\frac{\rho^2}{2}}\,.
	\end{align}
	Since $\psi_{\tau,B}^{D} \equiv 0$, this is always compatible with the IMBC. Clearly, $E_{\tau,0}=0$ [cf. equation~\eqref{eq:eigenvalues1}].

\section{Results and discussion}
\label{results}
\subsection{Isolated quantum dot}
\label{iso-dot}
\subsubsection{Continuum model}
\begin{figure}[t!]
	\centering	
	\includegraphics[width=.99\linewidth]{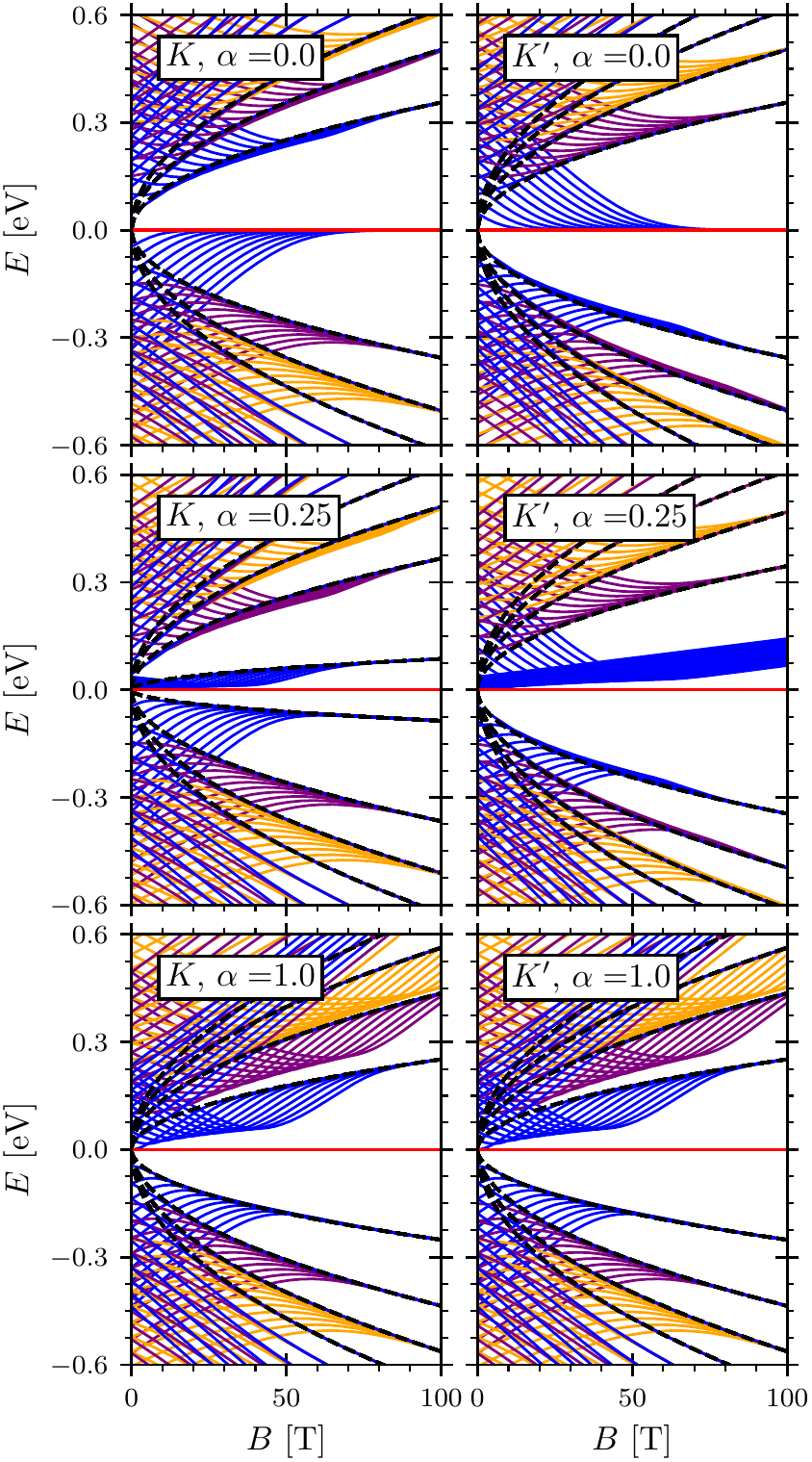}\\
	\caption{Eigenvalue spectra of an $\alpha-\mathcal{T}_3$ dot with radius $R=20$~nm. Solid lines give the solutions of~\eqref{eq:RB_dot} as a function of the perpendicular magnetic field $B$ in valleys $K$ (left) and $K'$ (right) when $\alpha=0$, 0.25, and 1 (top to bottom). Only results with \hf{$n_\rho=1$ (blue), 2 (violet) and 3 (orange) with} $-10\leq m \leq 10$ are shown. Flat bands are marked in red. Dashed black lines give the Landau levels~\eqref{eq:LL}.}
	\label{fig2}
\end{figure}

Figure~\ref{fig2} presents the analytical results for the magnetic field dependence of the energy spectra of (isolated) $\alpha-\mathcal{T}_3$ quantum dots with IMBC.  For all $\alpha$, we observe flat bands at $E=0$ (red lines) and a merging of the quantum dot states to the Landau levels characterised by quantum number $n_\tau$ (dotted curves) when the magnetic field increases. Note that $n_\tau=n_\tau (n_\rho,m)$ (the data show the results for $n_\rho\leq 3$  and $|m|\leq 10$). Different from normal semiconductors, the Landau levels exhibit a square-root dependence on $B$ [cf. equation~\eqref{eq:LL}], i.e., they are not equidistant. 

In the {\it graphene-lattice} model  ($\alpha=0$, top panels), we arrive at the same conclusions as previous work~\cite{GZCTFP11,TP17},  also for larger total angular and radial quantum numbers. According to the IMBC, the spectra show a broken particle-hole symmetry and $E_m\neq E_{-m}$, even for $B=0$ [where the eigenvalues are twofold degenerate ($E_\tau=E_{-\tau})$]. For $B>0$ time-reversal symmetry is broken and we have $E_\tau=-E_{-\tau}$. Combining the spectra of both valleys $K$ and $K'$, the symmetry is restored.

In the {\it $\alpha-\mathcal{T}_3$-lattice} model with $0 <\alpha < 1$ (see middle panels), the situation is the same for $B=0$, i.e., we find $E_m\neq E_{-m}$ and valley degeneracy $E_\tau=E_{-\tau}$. Clearly time-reversal symmetry is broken  at $B>0$ , but now $E_\tau\neq -E_{-\tau}$. As a consequence, the eigenvalues vary differently when $B$ is increased. Such valley-anisotropy has been found in the magneto-optical properties of (zigzag) $\alpha-\mathcal{T}_3$ nanoribbons~\cite{CXWLM19}. 

For the {\it Dice-lattice} model ($\alpha=1$, bottom panels), we have a specific situation. Here, $E_m =E_{-m}$ at $B=0$,  i.e., the state is now fourfold degenerate. When $B>0$ the states in each valley are still two-fold degenerate (Kramers degeneracy), and  the magnetic-field dependence of the energy spectrum is the same at the $K$ and $K'$ points.

\begin{figure}[t!]
	\centering
	\includegraphics[width=.99\linewidth]{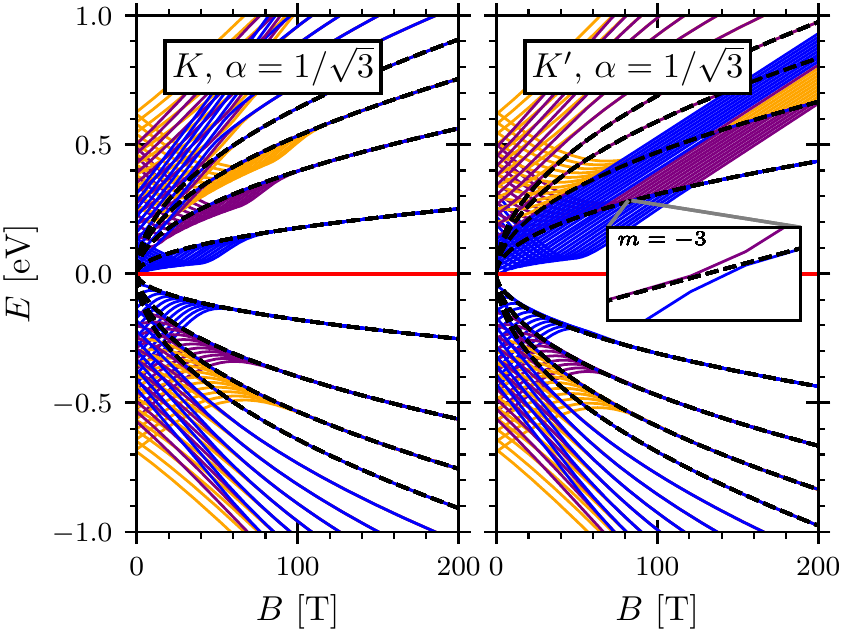}\\
		\caption{Eigenvalue spectra of a quantum dot with $\alpha=1/\sqrt{3}$ up to $B=200$~T (other model parameters and notation as in Fig. \ref{fig2}). Inset: Magnification of solutions with $m=-3$, and $n_\rho= 1$ (blue) respectively $n_\rho= 2$ (violet), at an avoiding crossing.}
	\label{fig3}
\end{figure}

Let us now discuss the convergence of the eigenvalues against the Landau levels in some more detail. The first Landau level  comprises all eigenvalues with  $m<0$; the higher Landau levels have contributions with $m<n_\tau$.  This holds for $K$ and $K'$, independent of $\alpha$. Obviously, the eigenvalues with positive (negative) energies cross the Landau levels first, before they converge towards these values from below (above) at the $K$ ($K'$) point when the magnetic field increases. The greater $\alpha$, the more pronounced this  kind of ``overshooting'' appears to be. This effect (being largest at $\alpha=1$) is not observed for negative (positive) energies  at $K$ ($K'$).  \hf{We note that in certain cases the eigenvalue levels  form a wide band of states and can be hardly resolved after bending up. In addition, looking for instance  at the blue curves for $\alpha=0.25$ ($K'$ point, $E>0$), it seems} that there is no convergence of this array of curves to a Landau level.  Figure~\ref{fig3} (right panel) shows, however,  that \hf{convergence of the $K'$-eigenvalue sets is reached  only for larger values of the} magnetic field. The inset demonstrates an avoided crossing for $m=-3$: While the eigenvalue belonging to $n_\rho=1$ (blue curve) converges to the first Landau level, the eigenvalue with $n_\rho=2$ (violet curve) tends to the second one. \hf{The same happens for the curves with other values of $m$.}

\begin{figure}[!]
	\centering
	\includegraphics[width=0.9\linewidth]{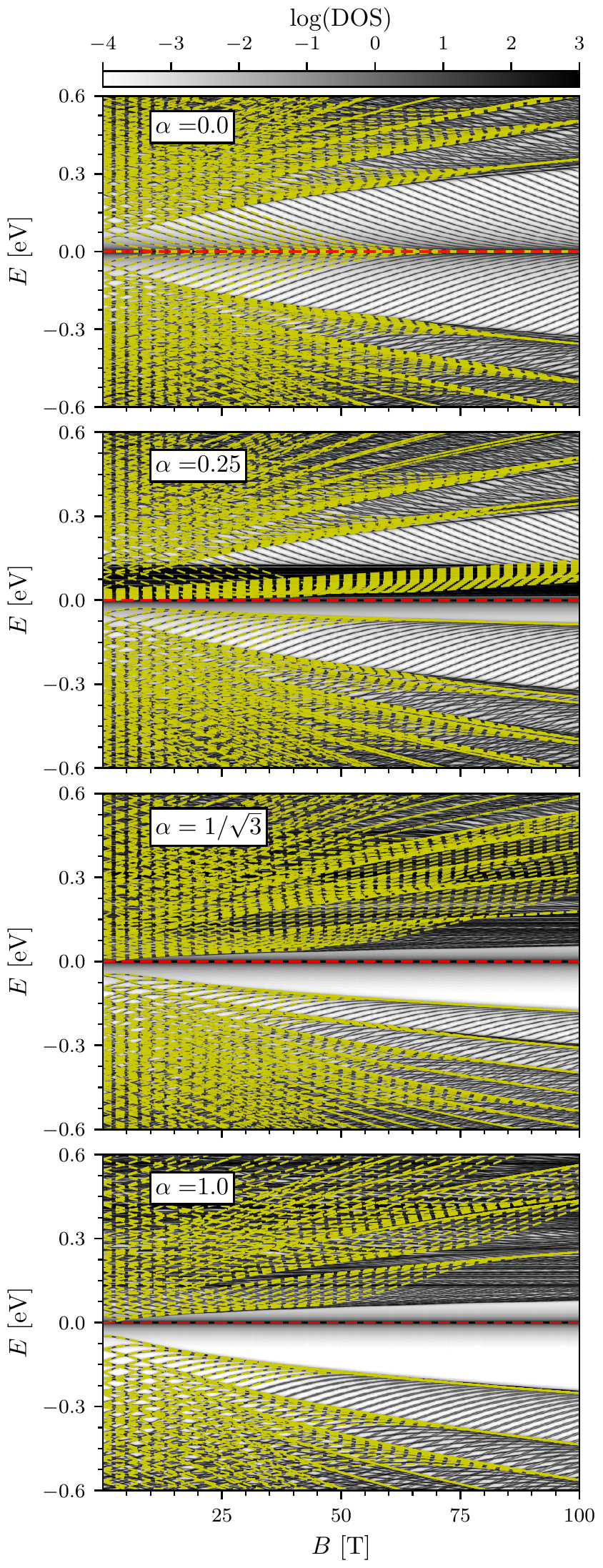}
	\caption{Logarithmic density of states, $\log({\rm DOS})$ (grey curves), of an $\alpha-\mathcal{T}_3$ quantum dot ($R=20$~nm) embedded in a circular ring-barrier potential $\Delta/t=0.8$ (width $W=5$~nm).  For comparison, the continuum model eigenvalues of Figs.~\ref{fig2} and~\ref{fig3} are  incorporated (yellow curves). Again the flat band is marked in red.}
	\label{fig4}
\end{figure} 
 
\begin{figure}[!]
	\centering
	\includegraphics[width=1\linewidth]{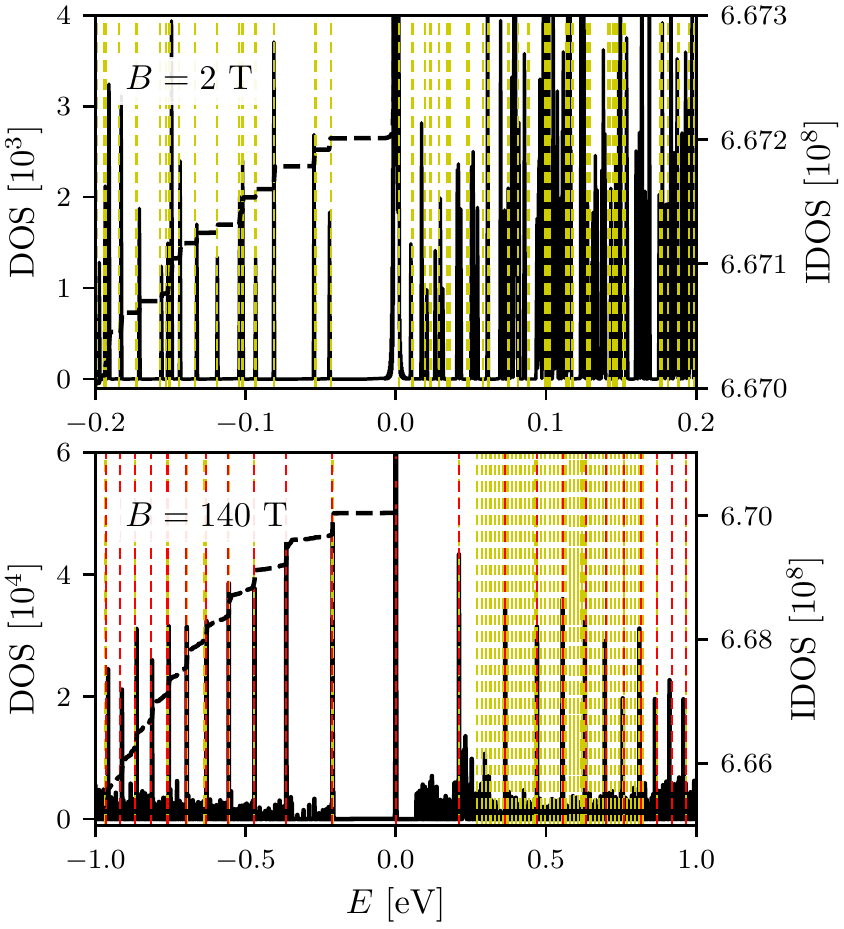}
		\caption{DOS (black lines, left axis) and integrated DOS (dashed lines, right axis; included for $E\leq 0$ only) of an $\alpha-\mathcal{T}_3$ quantum dot (where $\alpha=1/\sqrt{3}$) in a perpendicular magnetic fields: $B=2$~T (upper panel) and $B=140$~T (lower panel).  Yellow vertical lines mark the energy eigenvalues of the continuum model with IMBC. For $B=140$ T, the Landau levels are included (red lines). Other model parameters are as in Fig. \ref{fig4}.}
	\label{fig5}
\end{figure}  

\subsubsection{Tight-binding model}
We now analyse the validity range of the continuum model derived in the low-energy charge carrier regime close to the Dirac points  $K$ and $K'$. For this we consider the case of a circular dot imprinted on  the $\alpha-\mathcal{T}_3$ lattice, whereby the dot region is not surrounded by an infinite mass medium but by a ring (of width $W$ with finite mass potential $\Delta$, cf. Fig.~\ref{fig1}), which  has the same lattice structure as $D$. In this way particularly good result can be achieved if $\Delta/t > a/W$.  The eigenvalue problem of such a finite (non-interacting) system can be solved numerically, e.g., in a very efficient way by using the kernel polynomial method~\cite{WWAF06}. By the  kernel polynomial method we have also direct access to the local (L) density of states (DOS),
\begin{align}
\text{LDOS}(E)_i = \sum \limits_l |\langle i|l\rangle|^2 \delta(E - E_l)
\label{eq:ldos}
\end{align}
($i$ is a singled out lattice site and $n$ numbers the single-particle eigenvalues),
the DOS
\begin{align}
\text{DOS}(E) = \sum \limits_n \delta(E-E_n)\;,
\label{eq:dos}
\end{align}
and the integrated (I) DOS
\begin{align}
\text{IDOS}(E) = \int \limits_{-\infty}^E \text{DOS}(E')\mathrm{d}E'\,.
\label{eq:idos}
\end{align}

Figure~\ref{fig4} contrasts the DOS of our quantum dot lattice model with the eigenvalues of the continuum model, in dependence on the strength of the applied magnetic field $B$, for different values of $\alpha$. In general we can say that the continuum model provides an excellent approximation to the exact data for negative energies, regardless of $B$ and $\alpha$. At this point let us emphasise once again that if we had used a negative $\Delta$, positive and negative energy results would change roles. Comparing the data, one has to remember  that the numerical exact tight-binding approach takes into account larger angular momenta ($m$) than our continuum model calculation; therefore additional eigenvalues will appear also for $E<0$.  In the case of graphene ($\alpha =0$), we obtain a very good agreement also for positive energies, even though some features, such as the anti-crossing of energy levels, are not reproduced in the continuum model~\cite{GZCTFP11}.  At finite $\alpha$ (and $E>0$), the greatest difference between the continuum and tight-binding model results is the horizontal ``band" of states at low energies, where the width of this band increases when $\alpha$ is growing. These states are mainly localised at the quantum dot's boundary (see below), and can be related to the sublattice-dependent potential $\Delta$ along $\partial D$. \hf{Similar ``anomalous'' in-gap states were also found in two-dimensional \mbox{pseudospin-1} Dirac insulators and have been attributed to the boundary between two regions with different flat-band positions in a gapped Dice-lattice system~\cite{XL20}.  The edge states in our system have the same origin: The position of the flat band is shifted by $\Delta$ when changing from region I to II.}

Figure~\ref{fig5} compares the DOS of the tight-binding quantum-dot model and the distribution of the eigenvalues in the continuum IMBC model (with $n_\rho\leq 3$, $|m|\leq 20$)  for weak and strong magnetic fields. The heights of the steps in the integrated DOS can be taken as measure of the spectral weight of the corresponding eigenstates, particularly with regard to the degeneracy of the levels (note that the IDOS is not drawn for $E> 0$ for display reasons). The figure shows once again that the main energy levels are extremely well approximated by the continuum IBMC model for $E<0$. The sector $E>0$ is reproduced less accurately, obviously there are many states which are not taken into account within the continuum approach. For weak magnetic fields ($B=2$~T, upper panel), the Landau levels are more difficult to identify.  For high magnetic fields  ($B=140$~T, lower panel), states with large angular quantum numbers $m$ contribute to each Landau level. Note that we have included in the figure series of states which are not yet converged for the $n_\rho$- and $m$-values used (vertical dashed yellow lines).
\begin{figure}[t]
	\centering
	\includegraphics[width=0.8\linewidth]{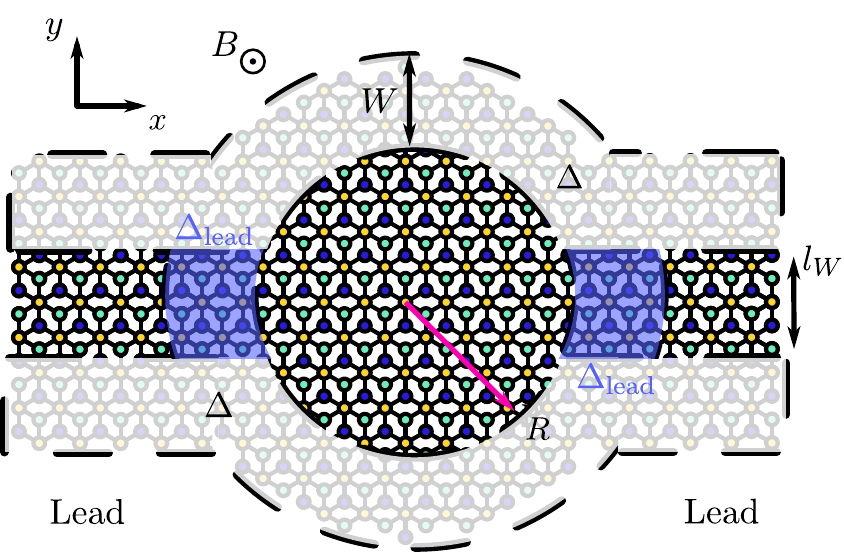}
	\caption{Drawing of the $\alpha-\mathcal{T}_3$ lattice quantum dot (radius $R$) contacted by leads (width $l_W$). The boundary condition is realised by a $W$-wide stripe with mass term $\Delta$ that covers the whole element. The leads are docked by an additional mass term $\Delta_{\rm lead}$ (blue region); the homogenous magnetic field $B$ points out of the plane. In the calculations we use $R=20$~nm, $\Delta/t=0.5$, $W=5$~nm, and $l_W =80 \sqrt{3}a - 2W$.}
	\label{fig6}
\end{figure} 

\subsection{Contacted quantum dot} 
\label{con-dot}
We now consider a more realistic situation, where the $\alpha-\mathcal{T}_3$ quantum dot is contacted by leads. The boundary of this ``device" is realised covering the whole setup by a sheath of width $W$ with a gapful band structure due to a (finite) 
mass term $\Delta$, see Fig.~\ref{fig6}.  To determine the conductance between the left (L) and right (R) leads in the limit of vanishing bias voltage, we employ the Landauer-B\"uttiker approach~\cite{Da95}:
\begin{align}
G = G_0 \sum \limits_{m\in\text{L}, n\in\text{R}} |S_{n,m}|^2
\label{conductance}
\end{align}
with  $G_0=2e^2/h$. $G_0$ is the maximum conductance per channel. The scattering matrix between all open (i.e., active) lead channels, $S_{n,m}$, can be easily calculated with the help of  the \texttt {Phyton}-based toolbox \texttt{Kwant}~\cite{GWAW14}.

\begin{figure}[!]
	\centering
	\includegraphics[width=1\linewidth]{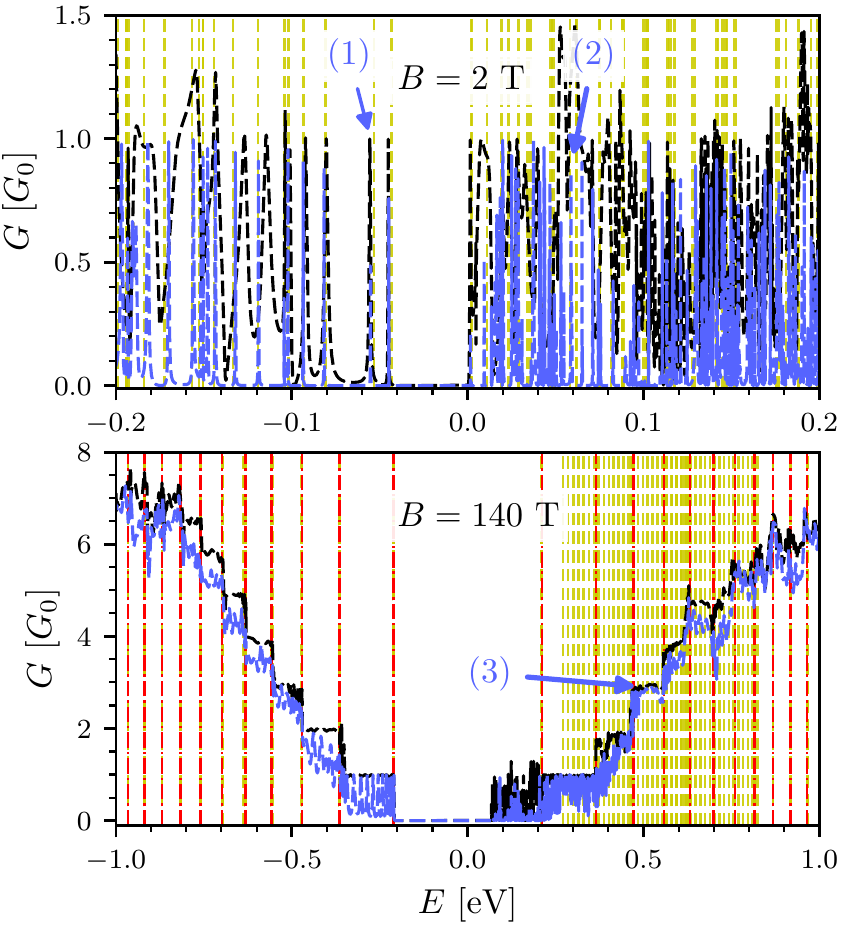}
	\caption{Conductance of the contacted $\alpha-\mathcal{T}_3$ quantum dot with $\alpha=1/\sqrt{3}$ as a function of energy for $B=2$~T (top) and $B=140$ T (bottom). The other dot parameters are as indicated in Fig.~\ref{fig6}. Results for $\Delta_\mathrm{lead}=0$ ($\Delta_\mathrm{lead}=0.2$~eV) are shown in black (blue). Yellow vertical lines are those included Fig.~\ref{fig5} as well.  Landau levels are marked by red lines. The LDOS for the selected signatures (1), (2) and (3) is given in Fig.~\ref{fig8} below.}
	\label{fig7}
\end{figure}

\begin{figure}[t]
	\centering
	\includegraphics[width=1\linewidth]{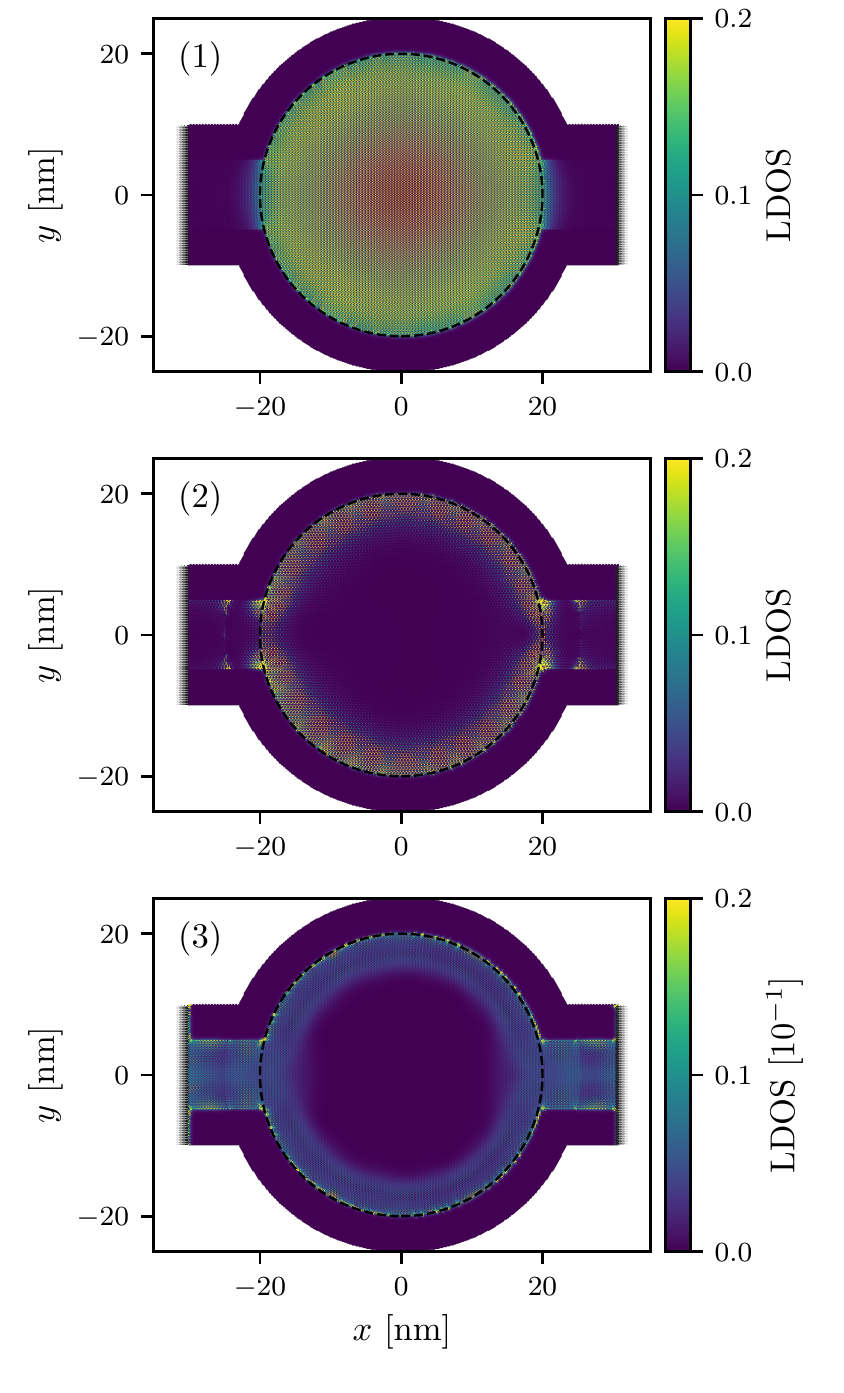}
	\caption{LDOS for the contacted $\alpha-\mathcal{T}_3$ quantum dot at the resonances indicated in Fig.~\ref{fig7} by (1), (2) [$B=2$~T; two upper panels] and~(3)  [$B=140$~T; lowest panel] for $\Delta_\mathrm{lead}=0.2$ eV.  Remaining parameters given in Fig. \ref{fig6}. The dashed line marks the dot boundary.}
	\label{fig8}
\end{figure}

\begin{figure}[t]
	\centering
	\includegraphics[width=1\linewidth]{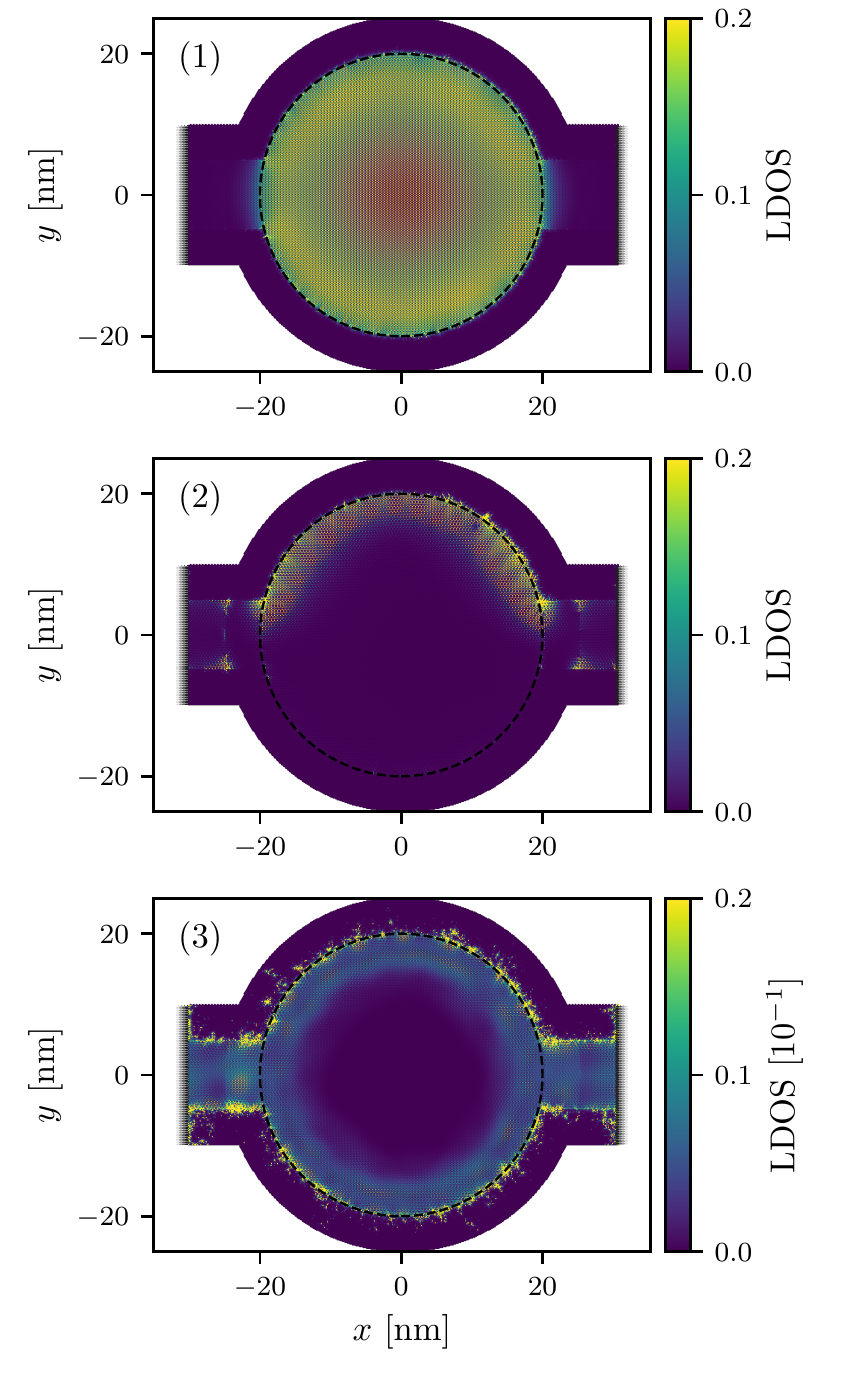}
	\caption{LDOS of the contacted $\alpha-\mathcal{T}_3$ quantum dot surrounded by a disordered circular ring. The LDOS is shown  for a single (but typical) realisation of the random mass term, where the  $\Delta_i$ are drawn out of the intervall~[0,1.6], i.e., $\hat{\Delta}=0.8$. Again we consider the resonances (1), (2) [$B=2$~T; two upper panels] and~(3)  [$B=140$~T; lowest panel] with system parameters as in  Figs.~\ref{fig6}, \ref{fig7} and~\ref{fig8}. 
	}
	\label{fig9}
\end{figure}

Figure~\ref{fig7} shows the conductance of the contacted $\alpha-\mathcal{T}_3$ quantum dot as a function of energy at weak (upper panel) and strong (lower panel)  magnetic fields. The conductance essentially probes the extended (current-carrying) states of the dot. Again, we  choose $\alpha=1/\sqrt{3}$,  in order to allow for a direct comparison with the DOS data of the isolated dot depicted in Fig~\ref{fig5}. Let us first consider the case $\Delta_{\rm lead}=0$ (black dashed lines). 
For  $B=2$~T,  we see that  the first five peaks at $E<0$ can be assigned to the eigenvalues of the continuum model for the isolated dot. For larger negative energies the conductance resonances will start to overlap, resulting in broader peaks, more specifically bands. In this range the rotation symmetry is completely destroyed by the contacts, and $m$ is not a good quantum number anymore.  For positive energies we recognise larger deviations from the continuum eigenvalues as is the case for the DOS (cf. Fig.~\ref{fig5}); overall much more conductive channels appear. At $B=140$~T, we observe the expected Landau level quantisation of the conductance. Obviously,  the steps respectively plateaus are less pronounced at positive and larger absolute values of the energy once again. The conductance quantisation basically breaks down if the cyclotron diameter $d_{c}=2|E|/v_{\rm F}eB$ exceeds the lead width $l_W$; in this case the charge carriers, moving  on a cyclotron trajectory along the quantum dot circumference, will miss the way out at the right lead.  

Working with additional barriers at the lead contacts ($\Delta_{\rm lead}=0.2$~eV, blue dashed lines), the conductance resonances are sharpened to some extent. This is because the dot region now is more self-contained.  Of course, the transmission of the device is reduced in  total when the barrier becomes too high (we have backscattering effects and, disregarding Klein tunnelling, only evanescent particle waves will enter the dot region). 

Further information about the nature of the states belonging to specific resonances can be obtained from the LDOS. Figure~\ref{fig8} records and visualises the spatial variation of the LDOS at the (resonance) energies  $E=-0.055$~eV (1),  $E=0.059$~eV (2) and $E=0.5$  (3) for $B=2$~T and $B=140$~T, respectively. For (1), the LDOS is almost  rotationally symmetric  (owing to the leads there is some weak asymmetry) and has a maximum at the centre of the quantum dot. This is in accord with the corresponding continuum solution ($m=0$, $s n_\rho=-1$ and $\tau=-1$), which according to equation~\eqref{eq:wavefunction_B} has  no angle dependence. Resonances at higher energy, belonging to larger values of  $m$, will lead to more complicated LDOS pattern (not shown). For~(2), the LDOS is more or less localised at the boundary of the quantum dot, i.e., this resonance will not correspond to a bulk state as~(1). Note that we find almost the same conductances, $G/G_0\simeq 0.98$ (1) and   $G/G_0\simeq 1$ (2), which indicates that we have one perfect current carrying (bulk or edge) state.  In both cases, we observe some scattering and `localisation' effects at the edges of the (lead) mass barrier. At resonance~(3), the LDOS at the quantum dot boundary is also much larger than those in the bulk (although by a factor of ten smaller compared to to cases (1) and (2); note the different scale of the color bar). Regardless of this, $G/G_0\simeq 2.9$, i.e., we have almost three perfect transport channels. In this case we already entered the quantum Hall regime, where quantum Hall edge states evolve which differ in nature from the edge state~(2).

\subsection{Disorder effects}
\label{dis-dot}
As a matter of course, imperfections will strongly influence the transport through contacted Dirac-cone systems~\cite{SF12a,PSWF13,FWPF18}. This holds true  even up to the point of complete suppression, e.g.,  by Anderson localisation~\cite{An58}. Nevertheless most of these nanostructures appear to be conducting~\cite{SSF09,SSF10}, simply because the (Anderson) localisation length exceeds the device dimensions for weak disorder in one or two dimensions~\cite{SSF09,SSF10}. In our case, the disorder caused by the boundary of the quantum dot is of particular importance. To model these disorder effects, we let the mass term fluctuate in the circular ring of width $W$. More precisely, we assume $\Delta \to \Delta_i$ in equation~\eqref{eq:Tight-Binding}, where  $\Delta_i$ is evenly distributed in the interval $[\Delta-\hat{\Delta}, \Delta+\hat{\Delta}]$ with $\hat{\Delta}<\Delta$, i.e., $\hat{\Delta}>0$ measures the disorder strength.  \hf{We note that only suchlike short-range disorder causes intervalley scattering, and thus may lead to Anderson localisation~\cite{SA02}. This holds at least in the case  $\alpha=0$ (graphene) and within the Dirac approximation. Long-range disorder, on the other hand, gives rise to intravalley scattering which is not sufficient to localise the charge carriers~\cite{BTBB07}.} 

\begin{figure}[b!]
	\centering
	\includegraphics[width=1\linewidth]{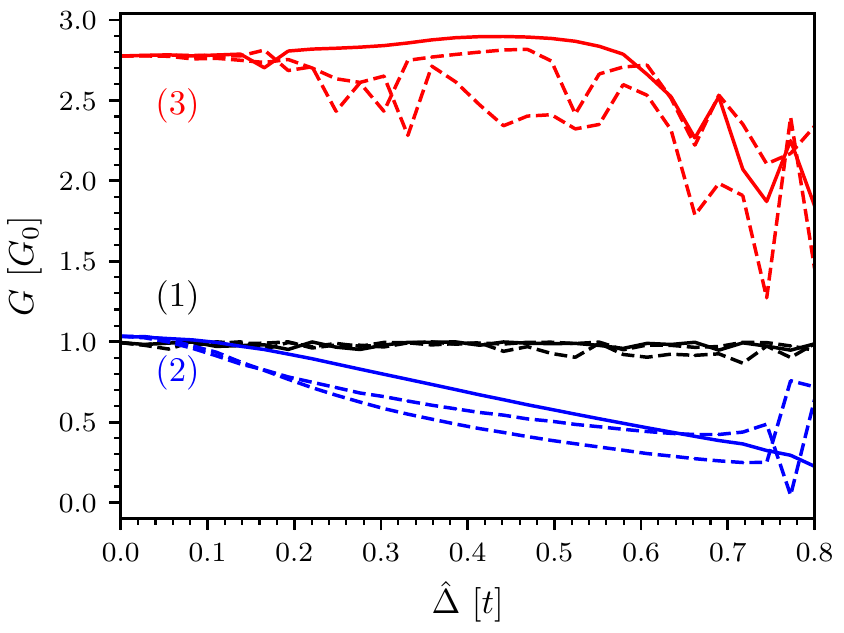}
	\caption{Conductance $G/G_0$ for the resonances (1) [black curves], (2) [blue curves] and (3) [red curves] (cf., Fig.~\ref{fig7}) calculated at different (discrete) disorder strengths $\hat{\Delta}$. Results obtained for the disorder realisation used in Fig.~\ref{fig9} (two other disorder realisations) are marked by solid (dashed) lines, which should guide  the viewer's eye only. All other parameters are as in the previous figures.  
	}
	\label{fig10}
\end{figure} 

Figure~\ref{fig9} illustrates how the LDOS shown in Fig.~\ref{fig8} for three characteristic resonances will change if we randomise  the mass potential $\Delta_i$ with strength $\hat{\Delta}=0.8$ in the ring covering the quantum dot. For this we have chosen  a randomly selected but from a physical perspective typical realisation (sample) and followed the resonances (1), (2) and (3) by increasing $\hat{\Delta}$ from zero to its final value 0.8.  Thereby the positions of the resonances (1) and (2) are slightly shifted compared to the ordered case: We find $E= -0.057$~eV (1) and $E=0.058$~eV (2) for the sample used in Fig.~\ref{fig9}. Since the plateau structure is completely destroyed for the (disordered) high-field case $B=140$~T, we will leave $E=0.5$~eV (3). It is obvious that the LDOS of the ``bulk-state'' resonance~(1) is not changed much by the edge disorder (upper panel). This is also reflected in the conductance $G/G_0(\hat{\Delta}=0.8)=0.93\simeq G/G_0(0)$.  A completely different behaviour is observed for the ``edge-state'' resonance (2). Here the LDOS is not homogeneously distributed along the periphery region anymore. Instead we find an imbalance between energy states (and associated transport channels) in the upper and lower half of the quantum dot, which depends on the specific sample of course. For other realisations the LDOS will be larger in the lower half of the quantum dot. In any case the conductance is substantially reduced, however, for example, we have  $G/G_0(\hat{\Delta}=0.8)=0.54$ for the depicted realisation. The  effect of the disorder is similarly strong for the quantum Hall edge-state resonance (3), $G/G_0(\hat{\Delta}=0.8)= 1.85)$, but here the LDOS is uniformly spread about the upper and lower halves of the quantum dot. Interestingly, it appears that now states can penetrate more deeply into the barrier region.

Finally, we show in Fig.~\ref{fig10} how the conductance depends on the disorder strength  $\hat{\Delta}$, for resonances (1), (2) and (3) and three different disorder realisations each. Despite the strong fluctuations at larger values of $\hat{\Delta}$, which clearly result from large local differences of the onsite energies and a varying overlap of energetically adjacent states, one observes a noticeable reduction of the conductance for the states (2) and (3) located primarily near the quantum dot boundary whereas the conductance of the bulk state (1) is only little affected. Since the spatial dimensions of the device are in the nanoscale regime, the conductance of our setup is not self-averaging. Determining the probability distribution for the LDOS and conductances from a large assembly of disorder realisations~\cite{SSBFV10} could be a promising approach to deal with this problem, but this is beyond the scope of the present work.

\section{Conclusions} \label{conclusions} To summarise, we considered  a generalisation of both 
graphene and Dice lattices, the so-called $\alpha-\mathcal{T}_3$ lattice, and studied the electronic properties 
of a quantum dot, imprinted on this material, in a perpendicular static magnetic field.  The quantum dot boundary condition was implemented in a consistent manner by an infinite mass term (circular ring having a finite band gap) in the continuum (tight-binding model) description.  For an isolated quantum dot we analysed the magnetic-field dependence of the eigenvalue spectra at the $K$ and $K'$ Dirac nodal  points and demonstrated significant  differences between the graphene, Dice and $\alpha-\mathcal{T}_3$ continuum model results, particularly with respect to the degeneracy and the convergence towards the Landau levels at high fields. The comparison of our analytical results with exact numerical data for the $\alpha-\mathcal{T}_3$ tight-binding lattice shows that the states with negative  band energies were generally satisfactory reproduced (if not too far away from the neutral point), whereas the lattice effects play a more prominent role at positive energies. For an contacted quantum dot, our transport calculations confirm the existence of transport channels, i.e., current carrying states, at weak magnetic fields, and Landau level quantisation of the conductance (related to quantum Hall edge states) at larger fields. The local density of states reveals the different physical nature of these states. The LDOS not only indicates how the boundary and the contacts affect the electronic structure, but also how disorder in the quantum dot's surrounding will influence its transport behaviour. While transport channels related to bulk resonances were less impacted, edge channel resonance and quantum hall edge states are strongly affected, giving rise to a  significant reduction of the conductance.

All in all, we are optimistic that the (strong) magneto-response of valley-contrasting quasiparticles in $\alpha-\mathcal{T}_3$ model materials  provides a good basis for promising valleytronics applications in near future.

\begin{acknowledgement}
The authors are grateful to R. L. Heinisch and C. Wurl for valuable discussions.
\end{acknowledgement}

\section*{Authors contribution statement}
Both authors contributed  equally to this work.

\bibliographystyle{epj}

\end{document}